\documentclass[aps,prb,twocolumn,amsmath,amssymb,superscriptaddress,nofootinbib]{revtex4-2}
\usepackage{graphicx}
\usepackage{dcolumn}
\usepackage{subfigure}
\usepackage{bm}
\usepackage{mathrsfs}
\usepackage{amsthm}
\usepackage{amsmath}
\usepackage{hyperref}
\usepackage{float}
\usepackage[utf8]{inputenc}

\hypersetup{
    colorlinks=true,
    linkcolor=blue,
    citecolor=blue,
    urlcolor=blue
}

\begin{document}

\title{Layer-selective and magnetic-field-enhanced transport of topological kink states in rhombohedral multilayer graphene}

\author{Chengyu Shen}
\affiliation{School of Physics and Technology, Nanjing Normal University, Nanjing 210023, People’s Republic of China}

\author{Zhe Hou}
\email{zhe.hou@nnu.edu.cn}
\affiliation{School of Physics and Technology, Nanjing Normal University, Nanjing 210023, People’s Republic of China}

\begin{abstract}
Topological valley kink states (VKSs), which are quantum valley Hall states emerging at the interfaces between adjacent domains with opposite valley Chern numbers, have attracted considerable interest in graphene-based systems. In this work, we investigate the quantum transport of VKSs in ABC-stacked rhombohedral multilayer graphene in the presence of Anderson disorder and a perpendicular magnetic field. Two prominent transport characteristics are revealed. First, in the absence of a magnetic field, the kink states exhibit strong layer polarization, with their wave functions predominantly localized and equally distributed on the outermost top and bottom layers. As a result, their transport properties are highly sensitive to the layer-selective disorder distribution. Second, under a perpendicular magnetic field, the layer-symmetric spatial distribution of VKSs is broken, leading to a significant reduction in the wave-function overlap between counter-propagating VKSs from opposite valleys. Consequently, intervalley scattering is suppressed, and the transmission of VKSs through disordered regions is substantially enhanced. Our results provide new insights into multichannel topological valley transport in rhombohedral multilayer graphene and demonstrates disorder-engineering and magnetic fields as effective approaches for manipulating the propagation of VKSs.

\end{abstract}
\maketitle

\section{Introduction}
Topological valley kink states (VKSs), which are topological valley Hall edge states residing at the interface of domains with opposite valley Chern numbers, have attracted growing interest in graphene systems owing to their flexible tunability and valley-dependent transport properties~\cite{Ju2015Topological, Yin2016Direct, Li2016Gate, Li2018A, Huang2024High, Martin2008, Jung2011Valley, Zhang2013PNAS, Semenoff2008, Zarenia2012, Cheng2016The}. Martin {\it et al}.~\cite{Martin2008} first proposed such one-dimensional (1D) kink states in voltage-biased bilayer graphene, where an inverted interlayer bias across the electric-field domain wall causes the valley Chern number to change from $1$ to $-1$ and induces two pairs of counterpropagating VKSs. In bilayer graphene, VKSs can also be realized at AB/BA stacking domain walls~\cite{Zhang2013PNAS}, where distinct stacking configurations produce opposite valley Chern numbers. In monolayer graphene, VKSs can be generated by introducing a staggered sublattice potential $\Delta \sigma_z$ that reverses its sign across a domain wall~\cite{Semenoff2008, Zarenia2012, Cheng2016The}. This sublattice potential opens a bandgap at the Dirac points and yields a valley Chern number of $\pm 1/2$ on either side of the domain wall~\cite{Xiao2007}, resulting in a single pair of VKSs. Recently, with the emergence of twisted bilayer graphene (TBG), it has been demonstrated that marginally TBG systems feature alternating AB/BA stacking domains~\cite{Yoo2019TBG, Ma2022Local}, in which VKSs inherently arise under a uniform perpendicular electric field~\cite{Yoo2019TBG, Ma2022Local, Prada2013Network, Efimkin2018Network, Huang2018mTBG, Xu2019Network, Beule2020ABTBG, Tsim2020Perfect, Fleischmann2020Perfect, Hou2024Arrays}. These VKSs have been shown to form 1D topological zigzag conducting channels~\cite{Tsim2020Perfect, Fleischmann2020Perfect} that yield quantized conductance plateaus~\cite{Hou2024Arrays}.

Compared to conventional topological states that are strictly localized at system boundaries, such as quantum Hall edge states~\cite{Klitzing1980New, Zhang2005Experimental}, quantum anomalous Hall (QAH) edge states~\cite{Haldane1988, Liu2008Quantum, Bestwick2015, Yu2010Quantized, Chang2013, Checkelsky2014}, and quantum spin Hall edge states~\cite{Kane2005Z2, Kane2005Quantum, Bernevig2006, Konig2007}, VKSs offer two distinct advantages. First, because topological VKSs reside at the interface between domains with contrasting valley Chern numbers, their spatial locations can be flexibly tuned by tailoring the domain wall geometries~\cite{Jiang2018Manipulation}. For instance, they can be engineered into network or junction configurations, enabling reconfigurable topological transport pathways and multi-channel device architectures~\cite{You2022Physical, Hou2020Metallic, Qiao2014Current, Li2016Gate, Li2018A, Huang2024High}. Second, these states are intrinsically coupled to the valley degree of freedom and exhibit prominent valley-dependent transport in graphene systems~\cite{Schaibley2016, Cheng2018Manipulation, Wang2021Topological}, holding significant promise for valley-filtering functional devices as well as valleytronics-based information encoding, storage, and processing.

\begin{figure}[t]
\centering
\includegraphics[width=1\linewidth]{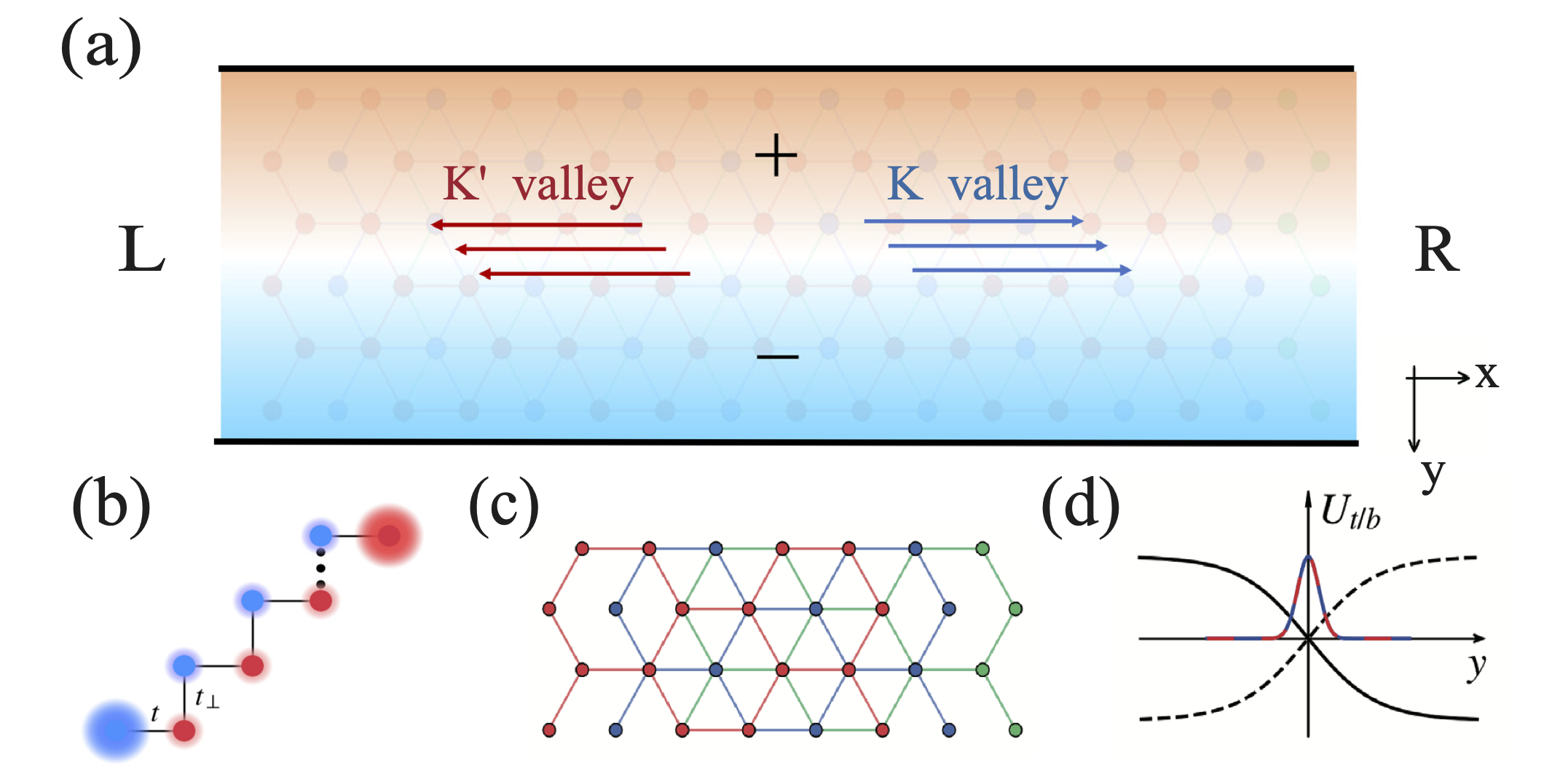}
\caption{
(a) Schematic of the rhombohedral multilayer graphene (RMG) transport device, with the inter-layer bias inversion (shown with the $\pm$ sign), electrodes (L/R), and valley-polarized VKSs labelled. (b) Side-view of the atomic structure of RMG, showing the layer-polarized distribution of wavefunctions of the VKSs, with red (blue) colors on sublattice A(B). The shadows denote the amplitude of wave functions of VKSs. (c) Schematic of an ABC-stacked trilayer graphene nanoribbon, where the three layers are distinguished by different colors. (d) Electrostatic potential $U_{t/b}$ for the top(bottom)-most layer, as functions of the transverse coordinate $y$, showing in a solid (dashed) curve. The out-of-plane electric field reverses its sign and a domain wall forms at the interface, leading to the emergence of VKSs localized at the domain wall (see the red and blue curves). 
} 
\label{fig_device}
\end{figure}

In recent years, ABC-stacked rhombohedral multilayer graphene (RMG) [see Figs.~\ref{fig_device}(b) and (c) for the atomic structures] has attracted significant attention owing to its highly gate-tunable band structure and a rich variety of correlated, topological, and superconducting phenomena. For instance, aligning rhombohedral trilayer graphene with hexagonal boron nitride (hBN) forms a moir{\'e} superlattice that reconstructs the low-energy electronic structure, giving rise to flat bands where electron-electron interactions dominate. A wide array of exotic quantum phases—including gate-tunable Mott insulating states~\cite{Chen2019Evidence}, superconductivity~\cite{Chen2019Signatures, Zhou2021Superconductivity}, correlated Chern insulating phases with ferromagnetism~\cite{Chen2020Tunable,Chen2022Correlated}, spin- and valley-polarized quarter metals~\cite{Zhou2021Half}, and intervalley coherent states~\cite{Arp2023Intervalley}—have been demonstrated in this system. Beyond trilayer graphene, rhombohedral tetralayer and pentalayer graphene host even richer phase diagrams due to their flatter bands. In rhombohedral tetralayer graphene, experiments have unveiled correlated Chern insulating states with a Chern number $C=4$~\cite{Sha2024Observation}, a layer-antiferromagnetic insulating state~\cite{Liu2023Spontaneous}, magnetic metallic states~\cite{Auerbach2025Isospin}, and superconductivity~\cite{Choi2025SC}. Similarly, in rhombohedral pentalayer graphene, tunable correlated Chern insulators with high Chern numbers ($C=5$)~\cite{Han2024Correlated,Han2024Large}, orbital multiferroicity~\cite{Han2023Orbital}, fractional QAH insulators~\cite{Lu2024Fractional}, signatures of chiral superconductivity~\cite{Han2025Signatures}, magnetic-field-enhanced superconductivity~\cite{Seo2026Family}, and the transdimensional anomalous Hall effect~\cite{Li2026Trans} have recently been observed.

Despite the aforementioned advances, a comprehensive theoretical understanding of VKSs in RMG remains lacking. The unique ABC stacking order, combined with the expanded layer degree of freedom, offers unprecedented avenues for controlling and harnessing topological VKSs. Furthermore, exploring the interplay between these topological edge channels, external electric and magnetic fields, and disorder is expected to uncover a rich spectrum of novel quantum transport phenomena.

In this work, we systematically investigate the quantum transport of VKSs in RMG under the impacts of disorder and a perpendicular magnetic field, focusing specifically on rhombohedral trilayer, tetralayer, and pentalayer graphene systems. By considering an RMG nanoribbon geometry [see Fig.~\ref{fig_device}(a)] subjected to a spatially reversed interlayer bias [Fig.~\ref{fig_device}(d)], we realize fully valley-polarized VKSs localized at the domain wall interface. Using the two-terminal device architecture illustrated in Fig.~\ref{fig_device}(a), two key transport features are revealed:

(i) In the absence of a magnetic field, the VKSs exhibit pronounced layer polarization, with their wave functions predominantly localized on the unbonded carbon atoms belonging to the outermost top and bottom layers. Moreover, the wave-function distributions are symmetric across the layers owing to the combined spatial inversion and time-reversal ($\mathcal{P}\mathcal{T}$) symmetry of the system. Upon introducing short-range Anderson disorder, intervalley backscattering is induced between the right-moving VKSs in the $K$ valley and the left-moving VKSs in the $K'$ valley [schematically indicated by blue and red arrows in Fig.~\ref{fig_device}(a)]. Notably, under layer-selective disorder, where impurity scattering is restricted to a single specific layer, the transport of VKSs displays a strong layer dependence: the conductance is heavily suppressed by disorder residing on the outermost top or bottom layers, whereas it remains robust against central-layer disorder.

(ii) More interestingly, applying a perpendicular magnetic field explicitly breaks the $\mathcal{P}\mathcal{T}$ symmetry, driving the wave functions of the $K$- and $K'$-valley VKSs to tilt toward the bottom and top layers, respectively. This spatial layer separation drastically reduces the spatial overlap between counterpropagating modes from opposite valleys, thereby strongly suppressing intervalley backscattering. Consequently, the transmission of VKSs across disordered regions is significantly enhanced, as confirmed by our numerical Landauer-B{\"u}ttiker conductance calculations.

These two prominent features are expected to hold generally for RMG systems with layer numbers $\mathcal{N} > 5$ due to the unique ABCAB... stacking order. Our work provides a microscopic understanding of multi-channel, topological valley-dependent transport in RMG systems, and establishes layer-selective disorder and perpendicular magnetic fields as effective control knobs for manipulating the propagation of VKSs.

The remainder of this paper is organized as follows. Section~\ref{sec_model} introduces the tight-binding model and the numerical method for transport calculations. Section~\ref{sec_trilayer} presents the main results including the band structure and conductance for rhombohedral trilayer graphene, which is taken as a protoprotical example. Section~\ref{sec_multilayer} extends the analysis to tetralayer and pentalayer graphene. Finally, Section~\ref{sec_conclusion} summarizes the main conclusions.

\section{Model and Method}
\label{sec_model}
We consider a ABC-stacked RMG in Fig.~\ref{fig_device}. The side view of this stacking configuration is illustrated in Fig.~\ref{fig_device}(b), where each layer is shifted relative to the adjacent one by a carbon--carbon bond length along the $x$-direction within the plane. A top-view of atomic-structure where the trilayer graphene is plotted as an example is shown in Fig.~\ref{fig_device}(c). An armchair edge termination has been adopted, so that a ribbon geometry with open boundary condition is formed. This choice is motivated by the fact that zigzag-edged ribbons generally host additional edge-localized states, which may coexist with VKSs and complicate the analysis ~\cite{Jung2011Valley, Nakada1996,Ryu2002Topological,Wassmann2008Stability,Okada2008Energetics}. In contrast, armchair-edged ribbons are free from such edge effect, allowing the VKSs to be clearly identified within the bulk gap. 

We employ a tight-binding model for rhombohedral $\mathcal{N}$-layer graphene under a perpendicular magnetic field. By including the nearest-neighbor intralayer and interlayer hoppings, the Hamiltonian reads 
\begin{equation}
\begin{aligned}
H = & \sum_{l,i} \epsilon_{l,i} c_{l,i}^{\dagger} c_{l,i}
+ t \sum_{l, \langle i,j \rangle} 
e^{i\phi_{ij}} c_{l,i}^{\dagger} c_{l,j} 
+  \\
& + t_{\perp} \sum_{l, \langle i, j \rangle} \left( c_{l,i}^{\dagger} c_{l+1, j} + \mathrm{H.c.} \right),
\end{aligned}
\label{eq_ham}
\end{equation}
where $c_{l,i}^{\dagger} (c_{l,i})$ is the creation (annihilation) operator at site $i$ in layer $l$, $\epsilon_{l, i}$ is the on-site potential, $t=2.6$ eV is the nearest-neighbor intralayer hopping integral, and $t_\perp= 0.4$ eV is the perpendicular interlayer hopping integral. 

The on-site potential consists of two terms: $\epsilon_{l, i} = V_l(y_i) + U_l({\bf r}_i)$. The first term denotes the electrostatic potential resulting from the reversed interlayer bias, and the second one denotes the Anderson disorder. The potential $V_l(y_i)$ is essential in constructing two adjacent domains with opposite valley Chern numbers~\cite{Zhang2013PNAS,Lee2016}. It depends on the layer index $l$ and the $y$-coordinate of the atomic site, but is uniform along the transport direction ($x$-direction). Its explicit form reads:
\begin{align}
V_l(y_i) = V_0  \tanh{\left( \frac{y_0 - y_i}{\lambda} \right) }  \left( l - \frac{\mathcal{N}+1}{2}   \right) 
\end{align}
for $ l = 1,2,\dots,\mathcal{N}$. Here a smooth $\tanh(x)$ function has been used for describing the $y$-dependence of the electrostatic potential, with $\lambda$ denoting the smoothness or the width of the electric-field reversal domain wall. To avoid an abrupt potential change that would induce strong intervalley scattering (see Appendix~\ref{app_inplaneProfile}), $\lambda$ is set to be much larger than the in-plane atomic distance. The layer-dependence is approximated by a linear function by assuming a constant potential difference $V_0$ between adjacent layers. Here the electrostatic screening effect has been omitted since it does not qulitatively changes our conclusions (see Appendix~\ref{app_layerProfile}). By setting $y_0 = 0$, i.e., the middle line of the nanoribbon, the system has inversion symmetry $\mathcal{P}$ in the absence of disorder. 

Due to the presence of the electric-field domain wall, linearly-dispersing VKSs emerge within the bulk band gap, with their number determined from the topological valley Chern number. For an $\mathcal{N}$-layer rhombohedral graphene, within the low-energy continuum description, the valley Chern number is calculated as (see details in Appendix~\ref{app_valleyChern})
\begin{equation}
C_{\xi} = \frac{\mathcal{N}}{2} \xi \mathrm{sgn}(\Delta),
\label{eq_valleyChern}
\end{equation}
where $\xi =\pm1$ is the valley index denoting the $K$($K'$) valley, and $\Delta$ denotes the potential difference between the top-most and bottom-most layers, with $|\Delta| = ( \mathcal{N}-1) V_0$ in our case ($V_0>0$). When $\Delta$ changes sign across the domain wall, the valley Chern number changes by $\mathcal{N}$, leading to the emergence of $\mathcal{N}$ VKSs for each valley. Besides, the VKSs in different valleys propagate in opposite directions due to the opposite helicities.

The Anderson disorder introduced into the on-site term takes the uniform box distribution:
\begin{align}
U_l({\bf r}_i) \in  [-W_l/2, W_l/2],
\end{align}
where $W_l = \eta_l W$, with $\eta_l = 1$ or 0 denoting the presence (absence) of disorder in the $l$-th layer, and $W$ characterizing the disorder strength.

The effect of a perpendicular magnetic field is incorporated by the Peierls substitution, with a phase
\begin{equation}
\phi_{ij} = - \frac{e}{\hbar} \int_{\mathbf{r}_i}^{\mathbf{r}_j} \mathbf{A} \cdot d\mathbf{l} 
\end{equation}
included into the intralayer hopping term. We adopt the Landau gauge here: ${\bf A} = (-By, 0, 0)$. This gauge choice does not break the $\mathcal{P}$ symmetry. Besides, it does not break the crystalline translational symmetry in the $x$-direction either, which allows us to calculate the band structure of the nanoribbon. 

To compute the transport properties, we divide the transport system in Fig.~\ref{fig_device}(a) into three parts: the left and right leads consisting of semi-infinite RMG nanoribbons, and the central scattering region, where disorder exits. Within the framework of the non-equilibrium Green's function method, the effect of  leads is incorporated via the self-energies ${\bf \Sigma}^r_{L}$ and ${\bf \Sigma}^r_{R}$. The retarded Green's function of the central region is given by ${\bf G}_c^{r}(E) = \left[ (E+ i0^+) {\bf I} - {\bf H}_C - {\bf \Sigma}^r_L(E) -  {\bf \Sigma}^r_R(E) \right]^{-1}$, where $E$ is the energy, $\bf I$ is the identity matrix, and ${\bf H}_C$ is the Hamiltonian matrix of the central region. By defining the linewidth functions ${\bf \Gamma}_{L,R} = i \left[ {\bf \Sigma}^r_{L,R} - ({\bf \Sigma}^r_{L,R})^{\dagger} \right]$ and the advanced Green's function ${\bf G}_c^a(E) = [{\bf G}^r_c(E)]^\dagger$, the two-terminal conductance at zero temperature is calculated as according to the Landauer formula~\cite{Datta1995,Meir1992,Jauho1994}
\begin{align} 
G = \frac{2e^2}{h} \mathrm{Tr} \left[ {\bf \Gamma}_L {\bf G}_c^r {\bf \Gamma}_R {\bf G}_c^a \right], 
\end{align} 
where the prefactor 2 accounts for the spin degeneracy. For convenience hereafter we define the conductance unit $G_0 \equiv 2e^2/h$ and omit it when plotting the $G$-curves.

\begin{figure*}[ht]
\centering
\includegraphics[width=0.98\textwidth]{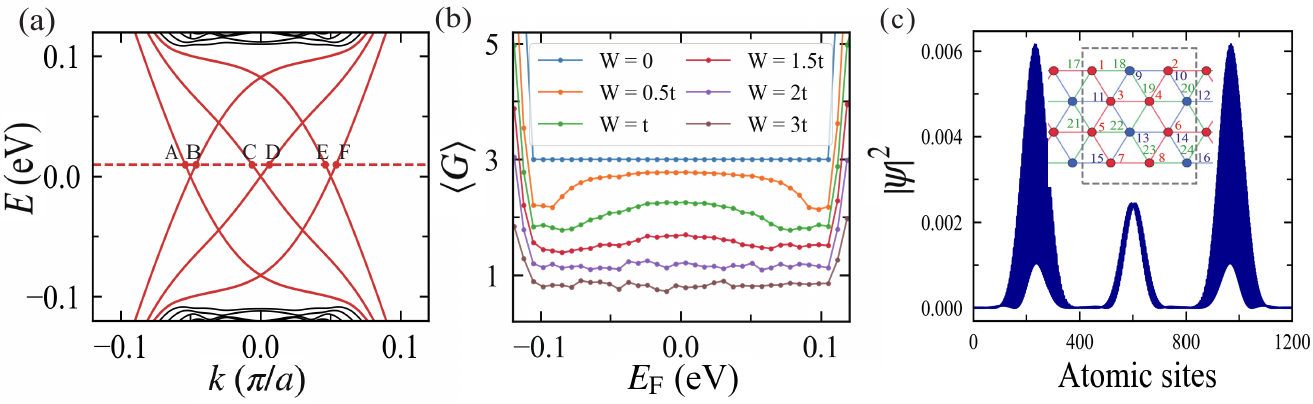}
\caption{ (a) Band structure of the trilayer graphene nanoribbon hosting the VKSs (shown in red curves). The nanoribbon width is $N=100$. 
(b) Conductance $\langle G \rangle$ as a function of the Fermi energy $E_F$ at different disorder strengths $W$. Here disorder exists in all-layers. 
(c) Wave function distributions of the VKSs, corresponding to point A in panel (a), where the Fermi energy $E_F=0.01$ eV. The inset shows the unit-cell of the nanoribbon of $N=2$, with the number denoting the indices of the atomic sites. }
\label{fig_tri_band}
\end{figure*}

In the numerical calculations, the nanoribbon geometry is defined as follows: each layer contains $2N$ atomic sites for each zigzag chain along the $y$-direction, and the central scattering region consists of $M$ unit cells [see the dashed box of the inset in Fig.~\ref{fig_tri_band}(c), where $N=2$ is shown] along the transport direction. The interlayer potential difference is fixed to $V_0 = 0.15$ eV unless otherwise stated. The magnetic field is characterized by the dimensionless flux per hexagon:$\phi \equiv \Phi/\Phi_0 = BS/\Phi_0$, where $S = (3\sqrt{3}/2)a^2$ is the area of a graphene unit hexagon with $a=0.142$ nm the in-plane carbon-carbon atomic distance, and $\Phi_0 = h/e$ the magnetic flux quantum. The length of the central region is fixed to $M=20$. For disorder calculations, the conductance is averaged over 50 configurations.

\section{Band Structure and Transport Properties of Rhombohedral Trilayer Graphene}
\label{sec_trilayer}
In this section, we first study the rhombohedral trilayer graphene by presenting its band structure and transport properties. In subsection A we discuss the results without magnetic fields, and in subsection B we include the effect of magnetic fields. The nanoribbon width is fixed to $N=100$, and the domain wall width is fixed to $\lambda=15 \sqrt{3} a /2$. The trilayer graphene plays as a protoprotical example and the conclusions drawn therein can be recovered in tetralayer and pentalayer graphene, as will be shown in Sec.~\ref{sec_multilayer}. 

\subsection{Results without magnetic fields}
Figure~\ref{fig_tri_band}(a) shows the band structure of a clean rhombohedral trilayer graphene nanoribbon. Three pairs of counter-propagating VKSs are identified (see the red curves), in accordance with layer layer number [see Eq.~(\ref{eq_valleyChern})]. The $K(K')$ valley VKSs have positive (negative) slopes in Fig.~\ref{fig_tri_band}(a), propagating rightward (leftward).

Figure~\ref{fig_tri_band}(b) shows the two-terminal conductance $G$ of the system. In the absence of disorder ($W=0$), the conductance exhibits integer values in units of $G_0$ as a result of the ballistic transport in the central region. For energies within the bulk gap ($|E|< 0.1$ eV), a quantized conductance plateau with value of $3G_0$ is observed, contributed from the three-pairs of VKSs. With the onset of disorder existing in all-layers, the ensemble-averaged conductance $\langle G \rangle$ gradually decreases with the disorder-strength $W$. Despite the topological origin of these channels, they are not robust against the short-range disorder since scattering between VKSs in opposite valleys happens due to their spatial overlapping. We note that there always exist a Kramer pair of VKSs related by the time-reversal symmetry $\mathcal{T}$, whose wave functions are fully overlapped in real space [see states A and F labeled in Fig.~\ref{fig_tri_band}(a) as an example, where the Fermi energy is fixed at $E_F=0.01$ eV].  We also note that when disorder is strong ($W \geq 2 t$), the propagation of VKSs are insensitive to $E_F$, with $\langle G \rangle$ exhibiting another plateaus within $|E_F| <  0.1$ eV.

To further characterize these VKSs, we analyze the spatial wave function distribution $|\psi|^2$ in Fig.~\ref{fig_tri_band}(c), where one representative state [state A in Fig.~\ref{fig_tri_band}(a)] is chosen for clarity. As expected, the wave function is localized at the domain wall where the out-of-plane electric field reverses its sign. More interestingly, it is distributed predominantly on the top and bottom layers, with weaker contribution from the middle layer, which is a consequence of the special stacking order of RMG~\cite{Zhou2023Layer}. Similar layer-selective distributions are found for the remaining states (B-F), indicating a general layer polarization of VKSs.

\begin{figure}[h]
\centering
\includegraphics[width=1.0\columnwidth]{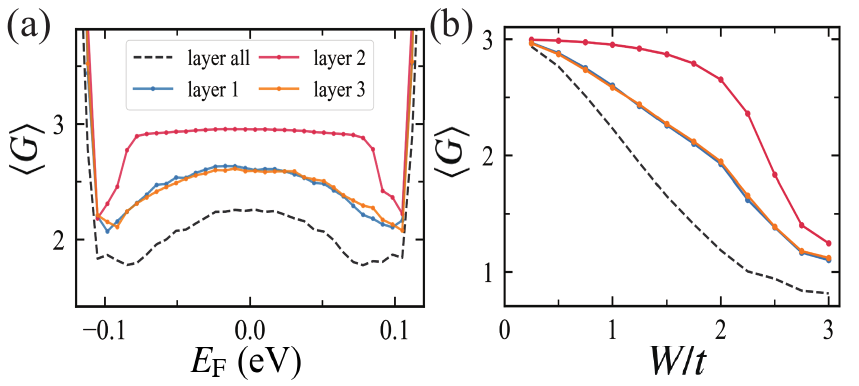}
\caption{ 
$\langle G \rangle $ as a function of $E_F$ at a fixed disorder strength $W=t$ in (a), and as a function of $W$ at a fixed Fermi energy $E_F = 0.01$ eV in (b). The four different curves correspond to disorder being introduced to all layers, or exclusively to layer 1, 2, and 3. 
}
\label{fig_tri_GzeroB}
\end{figure}

\begin{figure}[t]
\centering
\includegraphics[width=1.0\linewidth]{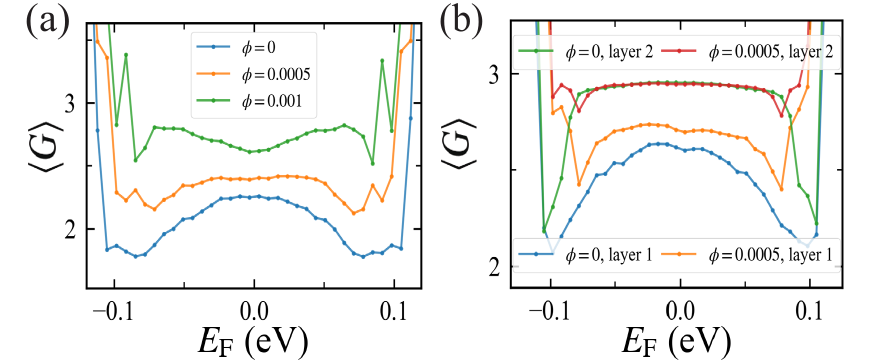}
\caption{
$\langle G \rangle$ as a function of $E_F$ at a fixed disorder strength $W=t$ in the absence ($\phi = 0$) and presence of a magnetic field for trilayer graphene. For (a) the disorder is applied to all layers, and for (b) the disorder is applied exclusively to layer-1 (the top-most layer) or layer-2 (the middle layer).
}
\label{fig_tri_GwithB}
\end{figure}

Motivated by the above layer-polarization property, we next consider the layer-selective disorder distribution, and study its influence on the conductance. In Fig.~\ref{fig_tri_GzeroB}(a) the disorder is applied exclusively to one layer with a fixed strength $W=t$. When disorder exists in the middle layer ($l=2$), the conductance remains relatively robust with its value close to the quantized one of $3G_0$. In contrast, when disorder is applied to either the top or bottom layer ($l=$1 or 3), a much stronger suppression on conductance is observed, with the two cases exhibiting nearly identical behavior. Besides, we also examine the conductance $\langle G \rangle$ as a function of disorder strength $W$ at a fixed Fermi energy $E_F=0.01$ eV, as shown in Fig.~\ref{fig_tri_GzeroB}(b). We find $\langle G \rangle$ decreases slower with $W$ in the middle-layer disorder case at $W\leq 2t$, and has values larger than the top/bottom layer case. These behaviors confirm that the dominant transport channels are primarily hosted by the outer layers, consistent with the wave function distribution in Fig.~\ref{fig_tri_band}(c). However, in the very strong disorder regime, the conductance is strongly suppressed regardless of the disorder location, indicating eventual destruction of all transport channels.

\begin{figure}[]
\centering
\includegraphics[width=1.0\linewidth]{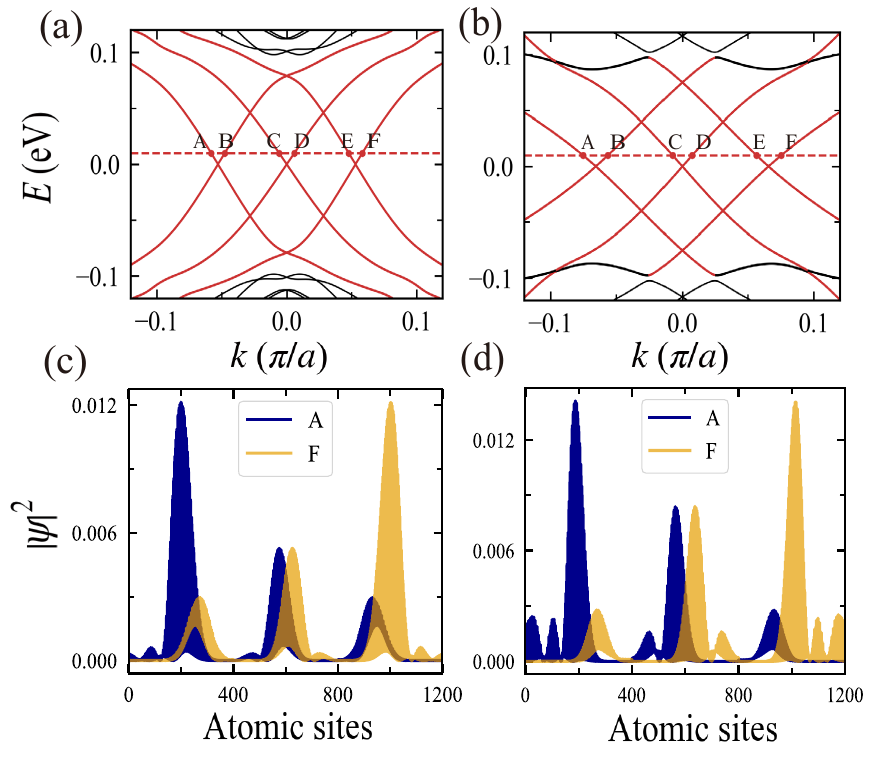}
\caption{
(a) and (b): Band structures of rhombohedral trilayer graphene at magnetic flux $\phi = 0.0005$ and $\phi = 0.001$, respectively. 
(c) and (d): Spatial distributions of wave functions at $E = 0.01$ eV, corresponding to states A and F in panel (a) and (b), respectively. }
\label{fig_tri_wfB}
\end{figure}

\begin{figure*}[t]
\centering
\includegraphics[width=0.95\linewidth]{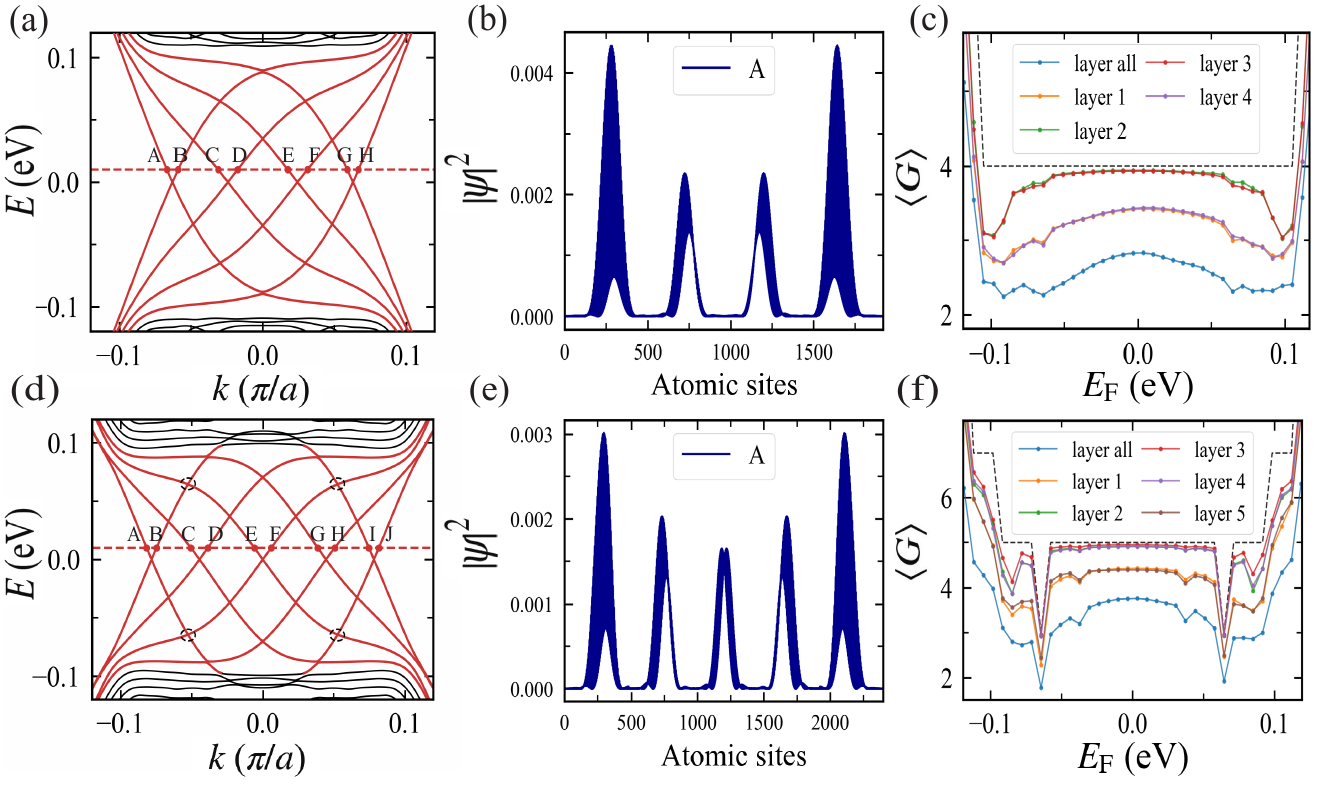}
\caption{
(a) and (d): Band structures of rhombohedral tetralayer and pentalayer graphene, respectively, hosting four and five pairs of VKSs (labeled in red curves). 
(b) and (e): Wave function distributions of VKSs of state A labeled in (a) and (d), respectively.
(c) and (f): $\langle G \rangle$ as a function of $E_F$ at a fixed disorder strength $W=t$. The disorder is applied to all layers (blue curves), or exclusively to one layer. The black dashed curves indicate the quantized conductance in the clean limit ($W=0$). 
}
\label{fig_fourfive_band}
\end{figure*}

\subsection{Magnetic Field Effects}
Next we include the effect of a perpendicular magnetic field by setting a nonzero $\phi$ in the tight-binding model. The magnetic field preserves the inversion symmetry $\mathcal{P}$ but breaks the time-reversal symmetry $\mathcal{T}$, and so the combined symmetry $\mathcal{P T}$ . The disorder can exist in all-layers, or exclusively in one-layer. In Fig.~\ref{fig_tri_GwithB}(a) we first discuss the all-layer disorder case. Within the bulk gap, we find that a weak magnetic field significantly enhances the conductance compared to the zero-field case, similar to the phenomenon reported in bilayer graphene~\cite{Wang2019Enhanced}. Besides, such enhancement becomes more prominent as $\phi$ increases (see the green curve of $\phi=0.001$). In Fig.~\ref{fig_tri_GwithB}(b) we discuss the single-layer disorder case, where the disorder is applied exclusively to the top ($l=1$) or middle layer ($l=2$). The transport behavior for disorder in the bottom layer ($l=3$) is similar to the top-layer case and is not shown here. For the top-layer disorder, again we find similar magnetic-field enhanced conductance behavior. For the middle-layer disorder, the magnetic field has little effects on $\langle G \rangle$. 

To elucidate the above magnetic-field enhanced transport phenomenon, we examine the band structure and wave functions of the VKSs at $\phi = 0.0005$, as shown in Fig.~\ref{fig_tri_wfB}. We find that although the bands above the bulk gap are largely altered by magnetic fields, the linearly-dispersing VKSs remain intact with the energy spectra getting little affected. Besides, the magnetic fields make the VKSs from the same valley repel from each other, with their bands becoming almost parallel with each other. On the contrary, their real-space wave functions are substantially changed by the magnetic field. In Fig.~\ref{fig_tri_wfB}(a) we set the Fermi energy $E_F= 0.01$ eV (see the dashed curve) and mark the six VKSs by A to F. For clarity we plot the density distribution $|\psi|^2$ of states A and F in Fig.~\ref{fig_tri_wfB}(b), which are related by the inversion-symmetry operation $\mathcal{P}$. We note that the magnetic field has two effects on the wave functions: First, it pushes the $K(K')$-valley VKSs down (up) in the $y$-direction by the Lorentz force; Second, and most importantly, it breaks the layer-symmetric distribution of the VKSs and tilts the $K (K')$-states towards the bottom(top) layer as a result of the $\mathcal{P} \mathcal{T}$ symmetry breaking. The overlap between the VKSs at opposite valleys thus gets siginficantly reduced, and as a result, the intervalley scattering for disorder existing in all-layers is greatly suppressed. The conductance of VKSs across disordered regions thus gets siginificantly enhanced by the magnetic field. Besides, the spatial separation of VKSs increases by increasing the magnetic field strength. These explanations are in perfect agreement with the numerical results in Fig.~\ref{fig_tri_GwithB}(a).

Such spatial separation can also be found for the top-most or bottom-most layer, due to the layer polarization of the VKSs. This corresponds to the enhanced conductance for disorder only in layer 1, as shown in Fig.~\ref{fig_tri_GwithB}(b). However, for the middle-layer, we find that the amplitude of wave functions increases, with the maximum of $|\psi|^2$ changing from 0.0025 to 0.005 [see Fig.~\ref{fig_tri_band}(c) and Fig.~\ref{fig_tri_wfB}(c) for comparison]. Such increase cancels the effect of spatial separation of VKSs in different valleys, and induces an almost unchanged overall spatial overlap of wavefunctions. This is in agreement with the unaffected conductance in the presence of middle-layer disorder [see Fig.~\ref{fig_tri_GwithB}(b)]. 

The above results demonstrate unambiguously that even a weak perpendicular magnetic field can impose strong modulation on the spatial distribution of the VKSs, thereby controlling their transport in multilayer graphene systems.

\section{Extension to tetralayer and pentalayer graphene}
\label{sec_multilayer}

\begin{figure*}[ht]
\centering
\includegraphics[width=0.95\linewidth]{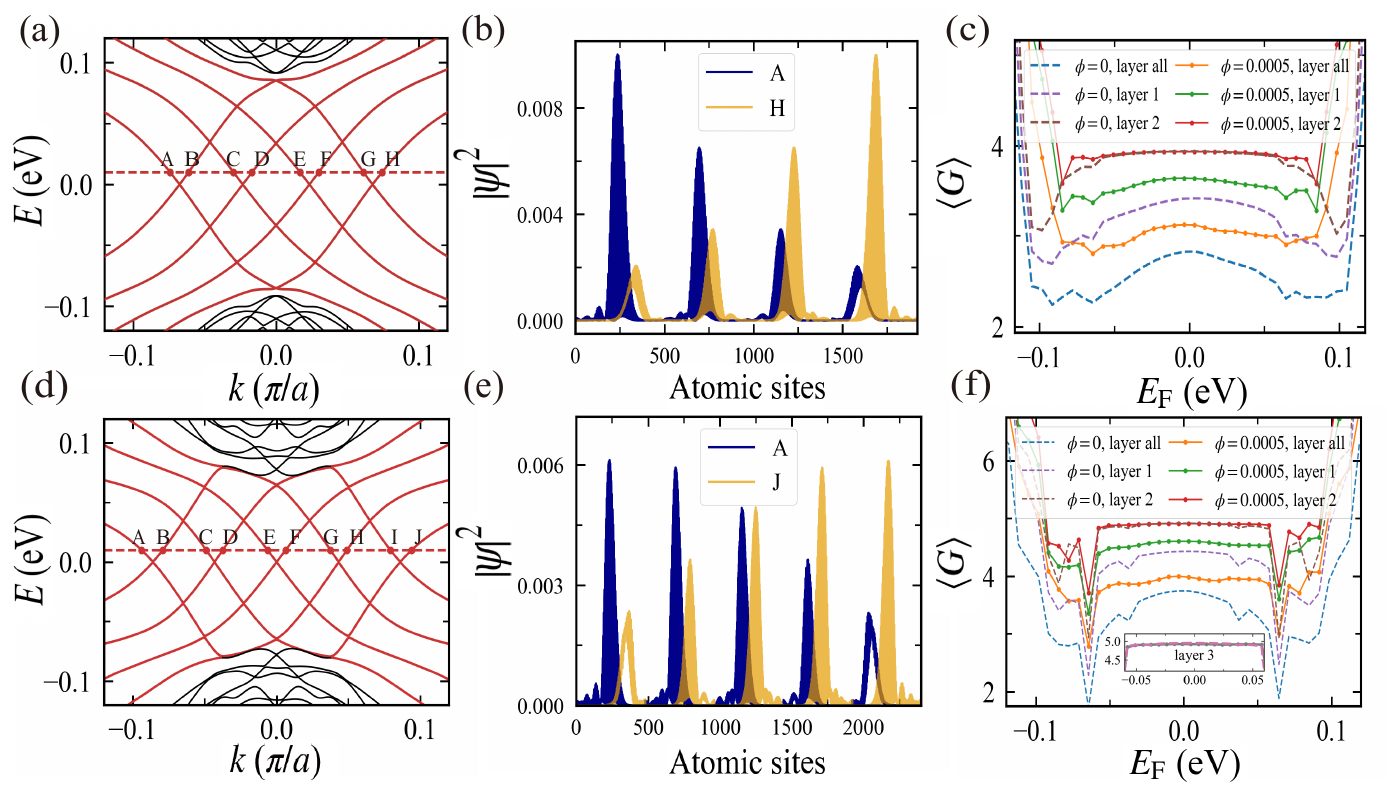}
\caption{
(a) and (d): Band structures of rhombohedral tetralayer and pentalayer graphene, respectively, under a perpendicular magnetic field. The VKSs are shown in red curves. The magnetic flux is set to $\phi = 0.0005$. The states at a Fermi energy $E_F=0.01$ eV are labeled by A to H for tetralayer, and A to J for pentalayer.
(b) and (e): Real-space wave function distributions of states labeled in (a) and (d), respectively. (c) and (f): $\langle G \rangle$ as a function of $E_F$ at a fixed disorder strength $W=t$. The disorder can exist in all layers , or exclusively in one layer. The inset shows the conductance curves with disorder exclusively in layer 3 with $\phi=0$ (dashed curve) and $\phi=0.0005$ (solid curve). 
}
\label{fig_fourfive_GwithB}
\end{figure*}

To further generalize our findings, we extend the analysis from trilayer graphene to rhombohedral tetralayer and pentalayer graphene, with ABCA and ABCAB stacking orders, respectively. Here, the domain wall width $\lambda$ is increased to $15\sqrt{3} a$ to suppress intervalley scattering of the VKSs as the layer number increases. Accordingly, we set the nanoribbon width to $N=120$ to avoid finite-size effects that would induce gap opening in the VKSs.

We first examine the system in the absence of a magnetic field. Figure~\ref{fig_fourfive_band} displays the results for tetralayer (upper panels) and pentalayer (lower panels) graphene. Figures~\ref{fig_fourfive_band}(a) and (d) present the band structures, where four and five pairs of counter-propagating, linearly dispersing VKSs emerge within the bulk gap, respectively. The corresponding wave functions for state A, shown in Figs.~\ref{fig_fourfive_band}(b) and (e), exhibit a highly layer-polarized structure; the wave function weight is predominantly localized on the outermost layers and systematically decays toward the inner layers. Due to $\mathcal{PT}$ symmetry, these wave function distributions remain symmetric across the layers. 

Figures~\ref{fig_fourfive_band}(c) and (f) show the conductance curves. In the clean limit ($W=0$), the conductance $G$ exhibits quantized plateaus at $4\,G_0$ and $5\,G_0$ for the tetralayer and pentalayer graphene, respectively, as indicated by the black dashed curves. Note that in Fig.~\ref{fig_fourfive_band}(f), the conductance plateau drops to $3\,G_0$ near $|E_F| \approx 0.0646$ eV due to residual intervalley scattering between the VKSs at the nanoribbon boundaries, which opens an anti-crossing gap in the band [see the black dashed circles in Fig.~\ref{fig_fourfive_band}(d)]. 

For tetralayer graphene, when disorder is introduced only to the outermost layers ($l=1$ and $l=4$), the conductance is severely decreased because of the large amplitude distribution of the VKSs. Conversely, when disorder is introduced to the inner layers ($l=2, 3$), scattering is significantly weaker, which allows the conductance to nearly recover its quantized plateau value. For comparison, we plot the conductance under all-layer disorder, which, as expected, yields the lowest conductance. A similar layer-selective transport behavior is observed in pentalayer graphene, where a clear conductance hierarchy corresponding to the spatial distribution of the disorder is resolved in Fig.~\ref{fig_fourfive_band}(f).

We next include the effect of a perpendicular magnetic field in Fig.~\ref{fig_fourfive_GwithB}. Figures~\ref{fig_fourfive_GwithB}(a) and (d) show the band structures at $\phi=0.0005$, with the corresponding wave functions of VKSs related by the inversion-symmetry shown in Figs.~\ref{fig_fourfive_GwithB}(b) and (e). Compared with the zero-magnetic-field case, we observe features similar to those in the trilayer system: while the bulk states are significantly altered by the magnetic field, the linearly dispersing VKSs remain intact. In contrast, the wave functions of these VKSs are significantly modified by the magnetic field. Specifically, the wave functions for the $K$ ($K'$) valley get pushed toward the bottom (top) layer as a result of the $\mathcal{PT}$ symmetry breaking induced by the magnetic field. Additionally, they shift in opposite directions along the $y$-axis, which further leads to a giant spatial separation of the VKSs between different valleys. Consequently, the spatial overlap of these VKSs is significantly reduced, suppressing the intervalley scattering and thereby enhancing VKSs transmission across the disordered regions.

Figures~\ref{fig_fourfive_GwithB}(c) and (f) present the corresponding transport results, plotting the average conductance $\langle G \rangle$ as a function of Fermi energy $E_F$. We first consider the case of all-layer disorder. For both tetralayer and pentalayer systems, the conductance is significantly enhanced compared to the zero-field scenario. A similar magnetic-field-induced enhancement is observed for top-layer-only disorder, which can likewise be attributed to the significantly reduced wave function overlap in the respective layers. When disorder exists in the central layers ($l=2$ for tetralayer and $l=2, 3$ for pentalayer), the conductance remains nearly unchanged, as evidenced in Figs.~\ref{fig_fourfive_GwithB}(c) and (f). This is again attributed to the increased amplitude of wave functions in the middle layer, which counteracts the effects of the spatial separation.

Overall, these results demonstrate that both the intrinsic layer polarization of the VKSs and the magnetic-field-enhanced transmission are universal features of RMG.

\section{Discussion and Conclusion}
\label{sec_conclusion}
Our results are expected to be applicable to other RMG systems with layer number $\mathcal{N} > 5$. First, the layer-polarization property should remain robust due to the characteristic rhombohedral stacking pattern, which gives rise to bonding and antibonding states, where two low-energy antibonding states have wave functions predominantly localized on the top and bottom layers. Consequently, the layer-selective response to disorder potentials is expected to be similar to that observed in our calculations ($\mathcal{N}=3, 4, 5$). Second, the application of a magnetic field breaks the $\mathcal{PT}$ symmetry and induces a layer imbalance of the VKSs, while states in opposite valleys exhibit opposite layer polarization due to inversion symmetry $\mathcal{P}$. As a result, the overlap between counter-propagating VKSs in different valleys is significantly reduced, leading to an anticipated magnetic-field-enhanced transport in the presence of disorder. Our work thus highlights the crucial role of symmetry breaking in tuning the valley-dependent transport properties in multilayer systems.

To conclude, we have systematically investigated quantum transport of VKSs in RMG using a tight-binding model and the non-equilibrium Green's function method. The VKSs exhibit pronounced layer polarization, with their wave functions predominantly localized on the outermost layers, making the transport highly sensitive to the layer distribution of disorder. Furthermore, a perpendicular magnetic field breaks the $\mathcal{PT}$ symmetry and reduces the overlap between counter-propagating modes in opposite valleys, thereby enhancing the transmission of VKSs in the presence of all-layer or outer-layer disorder. These findings provide a unified understanding of multichannel valley Hall transport in RMG and establish layer engineering and magnetic fields as effective tuning knobs for controlling topological valley transport.

\section*{Acknowledgments}
This work is supported by the National Natural Science Foundation of China under Grants No. 12304070. We are grateful to the Scientific Computing Center of Nanjing Normal University for doing the numerical calculations in this paper.

\section*{Data Availability} 
The simulation code that supports the findings of this article is available upon reasonable request.

\appendix

\section{Intervalley scattering induced by a sharp domain wall}
\label{app_inplaneProfile}
\begin{figure}[ht]
\centering
\includegraphics[width=0.8\linewidth]{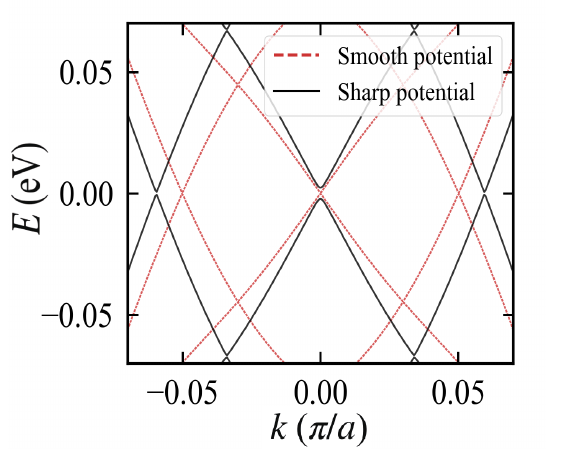}
\caption{
Comparison of band structures for different domain wall profiles for rhombohedral trilayer graphene nanoribbon. The solid curves represent the sharp potential profile where the interlayer bias changes abruptly (direct jump), while the dashed curves represent the smooth sigmoid function profile with $\lambda=15 \sqrt{3} a /2$. The nanoribbon width is $N=100$.
}
\label{fig_app1}
\end{figure}

To demonstrate that a sharp domain-wall profile induces strong intervalley scattering, we compare two types of domain wall potential profiles: a sharp, step-like profile and a smooth, sigmoidal profile. The corresponding band structures for rhombohedral trilayer graphene are shown in Fig.~\ref{fig_app1}, where solid curves denote the sharp profile and dashed curves correspond to the smooth case. For the sharp potential profile, prominent anticrossing gaps open at the band-crossing points between the VKSs of the $K$ and $K'$ valleys, signaling strong intervalley scattering and the breakdown of ideal VKSs~\cite{Ren2016Topological}. In contrast, a smooth potential profile drastically suppresses intervalley scattering by eliminating high-momentum Fourier components of the domain wall potential. Consequently, the band structure restores linear, gapless crossings that characterize well-defined zero-line modes.

\begin{figure}[ht]
\centering
\includegraphics[width=0.8\linewidth]{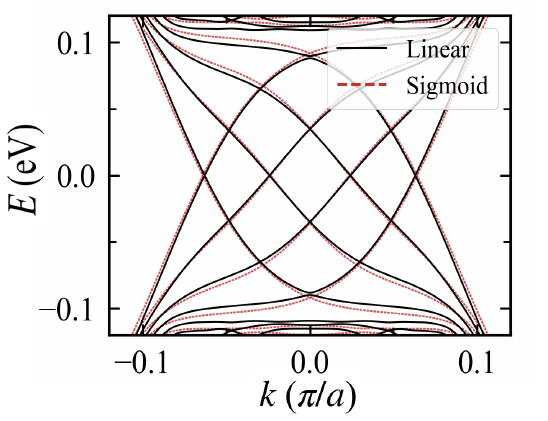}
\caption{
Comparison of band structures for different layer-dependent profiles for rhombohedral tetralayer graphene nanoribbon. The solid curves represent the linear potential with $V_1=0.225$ eV, $V_2=0.075$ eV, $V_3=-V_2$ and $V_4=-V_1$. The dashed curves represent the sigmoid potential with $V_1=0.225$ eV, $V_2=0.172$ eV, $V_3=-V_2$ and $V_4=-V_1$. Here the nanoribbon width is $N=120$. 
}
\label{fig_app2}
\end{figure}

\section{Discussion on the layer-dependence profile of the electrostatic potential}
\label{app_layerProfile}
In the main text, the electric field is modelled as a layer-dependent on-site potential with a linear profile across the layers. Away from the domain wall, the potential is $V_l(y_i) = \pm V_0  \left( l - \frac{\mathcal{N}+1}{2} \right)$, which corresponds to a uniform electric field across the multilayer system.

In realistic multilayer graphene, the electrostatic potential profile may deviate from a strictly linear form due to screening effects and charge redistribution among layers~\cite{Koshino2009,Koshino2010,Avetisyan2009}. Such effects can be incorporated in self-consistent Hartree or interaction-based approaches~\cite{Liu2025,Soejima2024,Zhang2019,Jung2013,Dong2024PRL,Guo2024FractionalChern}, which generally lead to a nonuniform potential distribution. However, the primary topological features studied in this work are determined by the sign structure of the interlayer potential difference rather than its detailed spatial profile. In Fig.~\ref{fig_app2} we adopt a nonuniform layer-dependent potential profile and plot its band structure for tetralayer graphene. Compared with the uniform profile (solid curves), the linearly-dispersing VKSs are almost unchanged. Besides, we have also checked that the wave functions remain quanlitatively unchanged. Therefore, we adopt the simplified linear potential profile in our setup. This approximation captures the essential physics relevant to kink states formation, disorder effects, and magnetic-field-induced transport enhancement.

\section{Calculation on the valley Chern number}
\label{app_valleyChern}
To characterize the topological properties of the RMG system, we compute its valley Chern number based on a low-energy continuum model. The effective Hamiltonian around the valley $\xi = \pm 1$ is written as
\begin{equation}
H_{\xi} = 
\begin{pmatrix}
V_1 & v_F\pi^\dagger & 0 & 0 & 0 & 0 & \cdots & 0 \\
v_F\pi & V_1 & t_\perp & 0 & 0 & 0 & \cdots & 0 \\
0 & t_\perp & V_2 & v_F\pi^\dagger & 0 & 0 & \cdots & 0 \\
0 & 0 & v_F\pi & V_2 & t_\perp & 0 & \cdots & 0 \\
0 & 0 & 0 & t_\perp & V_3 & v_F\pi^\dagger & \cdots & 0 \\
0 & 0 & 0 & 0 & v_F\pi & V_3 & \ddots & \vdots \\
\vdots & \vdots & \vdots & \vdots & \vdots & \ddots & \ddots & v_F\pi^\dagger \\
0 & 0 & 0 & 0 & 0 & \cdots & v_F\pi & V_{\mathcal{N}}
\end{pmatrix},
\end{equation}
where $\pi = \xi p_x + i p_y$ and $\pi^\dagger = \xi p_x - i p_y$, $v_F$ is the Fermi velocity, $t_\perp$ denotes the interlayer coupling, and $V_l$ is the layer-dependent electrostatic potential. This Hamiltonian is valid near the Dirac points and captures the essential low-energy physics of multilayer graphene under a perpendicular electric field.

For the $n$-th band with eigenvalue $E_{\xi n}(\mathbf{k})$ and Bloch eigenstate $|u_{\xi n}(\mathbf{k})\rangle$, the Berry connection is defined as~\cite{Fukui2005,Xiao2010}
\begin{equation}
\boldsymbol{\mathcal{A}}_{\xi n}(\mathbf{k}) = i \langle u_{\xi n}(\mathbf{k}) | \boldsymbol{\nabla}_{\mathbf{k}} | u_{\xi n}(\mathbf{k}) \rangle,
\end{equation}
and the corresponding Berry curvature is
\begin{equation}
\mathbf{\Omega}_{\xi n}(\mathbf{k}) ={\nabla}_{\mathbf{k}} \times \boldsymbol{\mathcal{A}}_{\xi n}(\mathbf{k}).
\end{equation}
The valley Chern number $C_{\xi}$ for a given valley $\xi$ ($\xi = \pm 1$ for the $K$ and $K'$ valleys, respectively) is obtained by integrating the $z$-component of the Berry curvature over a local momentum domain centered at that valley:
\begin{equation}
C_{\xi} = \frac{1}{2\pi} \sum_{n \in \mathrm{occ}} \iint \Omega_{\xi n}^z(\mathbf{k}) \, dk_x \, dk_y,
\end{equation}
where the summation runs over all occupied bands below the charge neutrality point.

In numerical calculations, $C_{\xi}$ is evaluated using the Wilson loop method~\cite{Fukui2005,Soluyanov2011,Soluyanov2012} on a discretized momentum-space grid encompassing valley $\xi$. Since the bands are isolated, we construct the Wilson loop for each occupied band individually. For a closed loop discretized into $N_w$ momentum points $\mathbf{k}_p$ ($p=1, \dots, N_w$ with $\mathbf{k}_{N_w+1} \equiv \mathbf{k}_1$), the Berry phase acquired by the $n$-th occupied band is extracted as the argument of the product of consecutive overlaps along the loop:
\begin{equation}
\Phi_{\xi n} = \operatorname{Arg} \left[ \prod_{p=1}^{N_w} \langle u_{\xi n}(\mathbf{k}_p) | u_{\xi n}(\mathbf{k}_{p+1}) \rangle \right].
\end{equation}
Summing the Berry phases over all occupied bands yields the net valley Chern number:
\begin{equation}
C_{\xi} = \frac{1}{2\pi} \sum_{n \in \mathrm{occ}} \Phi_{\xi n}.
\end{equation}
This approach provides a gauge-invariant and numerically stable scheme for computing topological invariant associated with each valley.

Numerically, we confirm that the valley Chern number for an $\mathcal{N}$-layer rhombohedral graphene system satisfies the general scaling relation
\begin{equation}
C_{\xi} = \frac{\mathcal{N}}{2} \, \xi \, \mathrm{sgn}(\Delta),
\end{equation}
as verified explicitly for $\mathcal{N} = 2, 3, 4, 5$ [see also Eq.~(\ref{eq_valleyChern}) in the main text], where $\Delta$ represents the interlayer potential bias across the system.


\end{document}